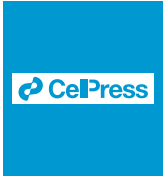
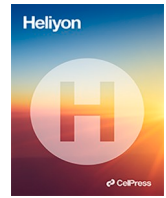

# A simple method for joint evaluation of skill in directional forecasts of multiple variables

Thitithep Sitthiyot [a,*], Kanyarat Holasut [b]

[a] *Department of Banking and Finance, Faculty of Commerce and Accountancy, Chulalongkorn University, Mahitaladhibesra Bld., 10th Fl., Phayathai Rd., Pathumwan, Bangkok, 10330, Thailand*
[b] *Department of Chemical Engineering, Faculty of Engineering, Khon Kaen University, Mittapap Rd., Muang District, Khon Kaen, 40002, Thailand*

ABSTRACT

Forecasts for key macroeconomic variables are almost always made simultaneously by the same organizations, presented together, and used together in policy analyses and decision-makings. It is therefore important to know whether the forecasters are skillful enough to forecast the future values of those variables. Here a method for joint evaluation of skill in directional forecasts of multiple variables is introduced. The method is simple to use and does not rely on complicated assumptions required by the conventional statistical methods for measuring accuracy of directional forecast. The data on GDP growth and inflation forecasts of three organizations from Thailand, namely, the Bank of Thailand, the Fiscal Policy Office, and the Office of the National Economic and Social Development Council as well as the actual data on GDP growth and inflation of Thailand between 2001 and 2021 are employed in order to demonstrate how the method could be used to evaluate the skills of forecasters in practice. The overall results indicate that these three organizations are somewhat skillful in forecasting the direction-of-changes of GDP growth and inflation when no band and a band of ± 1 standard deviation of the forecasted outcome are considered. However, when a band of ± 0.5% of the forecasted outcome is introduced, the skills in forecasting the direction-of-changes of GDP growth and inflation of these three organizations are, at best, little better than intelligent guess work.

## 1. Introduction

A forecaster is considered to have skill in directional forecast if the relative accuracy of a set of forecasts is better than the benchmark forecasts whose common choices are persistent forecast (*Up*, *Down*, or *No-change*) and random forecast which require no skill [1]. In economics and finance, this is generally known as measuring directional accuracy [2]. According to Sinclair et al. [3], the conventional statistical methods for evaluating directional accuracy of forecast are the Chi-square test, the Fisher's exact test, and the Pesaran-Timmermann [4] statistic. These statistical methods can be derived from the cell counts of forecasted/observed outcomes in the 2 × 2 contingency table. However, Sinclair et al. [3] and Pesaran and Timmermann [4] point out that, in order to use these statistical methods properly, various assumptions regarding the expected frequencies in the cell counts of forecasted/observed outcomes in the 2 × 2 contingency table, the continuity, the stationarity, and/or the types of distributions of forecasted/observed outcomes are required.

* Corresponding author.
*E-mail addresses:* thitithep@cbs.chula.ac.th (T. Sitthiyot), kanyarat@kku.ac.th (K. Holasut).

To avoid the problems associating with numerous assumptions required by the above-mentioned conventional statistical methods, an alternative is to use forecast skill scores. According to Sitthiyot and Holasut [5], forecast skill scores are widely used to assess the performance of forecast of binary outcomes/events in a number of scientific disciplines. Examples are [5–18]. Similar to the aforementioned conventional statistical methods for assessing accuracy of directional forecast, forecast skill scores for a single variable can be computed by using the cell counts of forecasted/observed outcomes in the 2 × 2 contingency table. However, they do not require assumptions about the expected frequencies in the cell counts of forecasted/observed outcomes in the 2 × 2 contingency table and various issues concerning the types of distributions of forecasted/observed outcomes.

Given that almost always forecasts for key macroeconomic variables such as growth of gross domestic product and inflation are simultaneously made by the same organizations, presented together, and used together in policy analyses and decision-makings [3], this study introduces a simple method for joint evaluation of skill in directional forecasts of multiple variables. Our method is constructed by applying the method for evaluating skill in forecasting binary events called the Prediction Skill Index (PSI) developed by Sitthiyot and Holasut [5] and the method for constructing a composite index devised by Sitthiyot and Holasut [19]. While there are many existing widely used forecast skill verification methods such as Peirce skill score (PSS) [20], Phi coefficient ($\varnothing$) [21], Heidke skill score (HSS) [22], and Clayton skill score (CSS) [23] which could be used in combination with Sitthiyot and Holasut [19]'s method for constructing a composite index in order to develop a method for joint evaluation of skill in directional forecasts of multiple variables, this study chooses PSI because it has a key advantage over the other conventional forecast skill verification methods, namely, PSS, $\varnothing$, HSS, and CSS in that it can distinguish the level of difficulty between the forecast of rare or extreme event and random event while these conventional forecast skill verification methods cannot and award the same score. For notations, let $a, b, c,$ and $d$ be the counts for *Hits*, *False alarms*, *Misses*, and *Correct rejections*. Let $n$ be the number of observations which equals $a + b + c + d$. The definitions and ranges of PSI, PSS, $\varnothing$, HSS, and CSS are shown in Table 1 whereas the advantage of PSI over PSS, $\varnothing$, HSS, and CSS in differentiating between the level of difficulty in forecasting rare or extreme event and random event are illustrated in Table 2.

The forecast skill scores for PSI, PSS, HSS, and CSS, as illustrated in Table 2, are from Sitthiyot and Holasut [5]. They are calculated based on the perfect forecast of rare or extreme event and that of random event using $n = 400$. Since Sitthiyot and Holasut [5] do not include $\varnothing$ in their analysis, this study computes the value of $\varnothing$ based on the same criteria employed for calculating PSI, PSS, HSS, and CSS as specified in Sitthiyot and Holasut [5]. The results indicate that PSI can distinguish the level of difficulty between the perfect forecast of rare or extreme event and random event by awarding the two events quite different scores while the other conventional skill verification methods, namely, PSS, $\varnothing$, HSS, and CSS cannot and award the two events the same score. Given the presence of rare or extreme event and random event which could have an impact on key macroeconomic variables, many of which are forecasted by the same organizations, presented together, and used together in policy analyses and decision-makings [3], our method could be used as an alternative tool to improve skills of forecasters who build models to forecast the direction-of-changes of multiple variables at the same time.

## 2. Materials and methods

### 2.1. PSI for directional forecast of a single variable

Let $i$ denote the total number of variables being forecasted, where $i = 1, 2, 3, ..., N$. As denoted in Introduction, let $a, b, c,$ and $d$ be the counts for *Hits*, *False alarms*, *Misses*, and *Correct rejections*, respectively. The number of observations ($n$) equals $a + b + c + d$. The 2 × 2 contingency table of two forecasted and two observed outcomes are shown as Table 3.

Given the cell counts for *Hits* ($a$), *False alarms* ($b$), *Misses* ($c$), and *Correct rejections* ($d$), as well as the total number of observations ($n$), PSI($i$) for directional forecast of a single variable or PSI(1) can be calculated as shown in equation (1).

**Table 1**
The definitions and ranges of PSI, PSS, $\varnothing$, HSS, and CSS.

| Methods | Definitions | Ranges |
|---|---|---|
| Prediction skill index (PSI) | $\dfrac{\left(\dfrac{\frac{a}{n} - \left[\left(\frac{a+b}{n}\right)\left(\frac{a+c}{n}\right)\right]}{\sqrt{\left(\frac{a+b}{n}\right)\left(\frac{a+c}{n}\right)}} + \dfrac{\frac{d}{n} - \left[\left(\frac{b+d}{n}\right)\left(\frac{c+d}{n}\right)\right]}{\sqrt{\left(\frac{b+d}{n}\right)\left(\frac{c+d}{n}\right)}}\right) - \left(\dfrac{\frac{b}{n} - \left[\left(\frac{a+b}{n}\right)\left(\frac{b+d}{n}\right)\right]}{\sqrt{\left(\frac{a+b}{n}\right)\left(\frac{b+d}{n}\right)}} + \dfrac{\frac{c}{n} - \left[\left(\frac{a+c}{n}\right)\left(\frac{c+d}{n}\right)\right]}{\sqrt{\left(\frac{a+c}{n}\right)\left(\frac{c+d}{n}\right)}}\right)}{2}$ | [-1, 1] |
| Peirce skill score (PSS) | $\dfrac{ad - bc}{(b+d)(a+c)}$ | [-1, 1] |
| Phi coefficient ($\varnothing$) | $\dfrac{ad - bc}{\sqrt{(a+b)(c+d)(a+c)(b+d)}}$ | [-1, 1] |
| Heidke skill score (HSS) | $\dfrac{a + d - \left[\frac{(a+b)(a+c)}{n}\right] - \left[\frac{(b+d)(c+d)}{n}\right]}{n - \left(\frac{(a+b)(a+c)}{n}\right) - \left[\frac{(b+d)(c+d)}{n}\right]}$ | [-1, 1] |
| Clayton skill score (CSS) | $\dfrac{a}{a+b} - \dfrac{c}{c+d}$ | [-1, 1] |

**Table 2**

The advantage of PSI over PSS, ∅, HSS, and CSS in differentiating between the level of difficulty in forecasting rare or extreme event and random event.

| Events | Forecasted/observed outcomes | | | | PSI | PSS | ∅ | HSS | CSS |
|---|---|---|---|---|---|---|---|---|---|
| | Yes/Yes (*a*) | Yes/No (*b*) | No/Yes (*c*) | No/No (*d*) | [-1, 1] | [-1, 1] | [-1, 1] | [-1, 1] | [-1, 1] |
| Rare or extreme | 1 | 0 | 0 | 399 | 0.550 | 1.000 | 1.000 | 1.000 | 1.000 |
| Random | 193 | 0 | 0 | 207 | 1.000 | 1.000 | 1.000 | 1.000 | 1.000 |

**Table 3**

The 2 × 2 contingency table of forecasted and observed outcomes.

| Forecasted outcomes | Observed outcomes | | Total |
|---|---|---|---|
| | Up | Down | |
| **Up** | *a* (Hits) | *b* (False alarms) | $a+b$ |
| **Down** | *c* (Misses) | *d* (Correct rejections) | $c+d$ |
| **Total** | $a+c$ | $b+d$ | $n = a+b+c+d$ |

$$\text{PSI}(1) = \frac{\left(\frac{\frac{a}{n}-\left[\left(\frac{a+b}{n}\right)\left(\frac{a+c}{n}\right)\right]}{\sqrt{\left(\frac{a+b}{n}\right)\left(\frac{a+c}{n}\right)}} + \frac{\frac{d}{n}-\left[\left(\frac{b+d}{n}\right)\left(\frac{c+d}{n}\right)\right]}{\sqrt{\left(\frac{b+d}{n}\right)\left(\frac{c+d}{n}\right)}}\right) - \left(\frac{\frac{b}{n}-\left[\left(\frac{a+b}{n}\right)\left(\frac{b+d}{n}\right)\right]}{\sqrt{\left(\frac{a+b}{n}\right)\left(\frac{b+d}{n}\right)}} + \frac{\frac{c}{n}-\left[\left(\frac{a+c}{n}\right)\left(\frac{c+d}{n}\right)\right]}{\sqrt{\left(\frac{a+c}{n}\right)\left(\frac{c+d}{n}\right)}}\right)}{2} \quad (1)$$

$-1 \leq \text{PSI}(1) \leq 1$

Note that the complete derivation of PSI(1) and the way in which it assesses skill in directional forecast of a single variable can be found in Sitthiyot and Holasut [5]. The value of PSI(1) is between −1 (the worst forecast skill) and 1 (the best forecast skill), with 0 being indifferent from the random forecast which is set to be the reference forecast. Like other forecast skill scores, PSI(1) measures skill of forecaster as a percentage improvement over the reference forecast, not in form of statistical significance. Let PSI(1), PSI(1)$^{\text{perf}}$, and PSI(1)$^{\text{ref}}$ denote the value of PSI calculated from the forecasted outcomes and the observed outcomes of a single variable, the value of PSI that would be achieved by perfect forecast of a single variable, and the value of PSI being consistent with the random forecast of a single variable, respectively. The skill score for PSI(1) can be computed as shown in equation (2).

$$\text{Skill score for PSI}(1) = \left(\frac{\text{PSI}(1) - \text{PSI}(1)^{\text{ref}}}{\text{PSI}(1)^{\text{perf}} - \text{PSI}(1)^{\text{ref}}}\right) * 100\% \quad (2)$$

If PSI(1) = PSI(1)$^{\text{perf}}$, the skill score for PSI(1) attains its maximum value of 100%. If PSI(1) = PSI(1)$^{\text{ref}}$, the skill score for PSI(1) = 0% which indicates that there is no improvement over the reference forecast. If the forecasted outcome being evaluated is inferior to the reference forecast based on the PSI method, the skill score for PSI(1) < 0%. Note that since PSI(1)$^{\text{perf}}$ = 1 and PSI(1)$^{\text{ref}}$ = 0, we can directly compare the values of PSI(1)s in order to assess whether skill scores of different forecasters computed based on the PSI method are different from each other. While the random forecast whose value is set to be 0 is used as the benchmark against which a forecast can be judged according to the PSI method, Sitthiyot and Holasut [5] show numerically that other benchmarks such as always forecasting *Up*, *Down*, or *No-change* can also be used since the PSI method gives the value of 0 in all three cases.

While the term 'directional accuracy' is usually employed in economics and finance when it comes to the issue of assessing quality of forecast, it is important to clearly distinguish the difference between forecast skill and forecast accuracy [5]. As discussed in Introduction, forecast skill is referred to the relative accuracy of a set of forecasts with respect to some sets of benchmark forecasts such as persistent forecast in the same direction and random forecast [1]. According to Sitthiyot and Holasut [5], a highly skilled forecaster would generally tend to have a high rate of forecast accuracy but the reverse is not necessarily true. It depends on the types of events/problems being forecasted, their levels of difficulty, and the current state of scientific knowledge used in forecasting such events/problems. For example, being able to correctly forecast today that tomorrow the sun will rise in the east would probably hit 100% forecast accuracy rate but this event hardly requires any skill. In contrast, forecasting today whether tomorrow there will be a financial crisis correctly would require extraordinary skills. Note that if we were to measure performance of an exceptionally high-skilled forecaster in forecasting these two events by using the accuracy measures such as mean squared error (MSE) and/or Theil inequality coefficient (U) [25] as widely used in economics and econometrics, there would be no difference in scores among these events. Given $\text{MSE} = \frac{\sum_{i=1}^{n}(A_i - F_i)^2}{n}$ and $U = \frac{[\sum_{i=1}^{n}(F_i - A_i)^2]^{\frac{1}{2}}}{[\sum_{i=1}^{n} A_i^2]^{\frac{1}{2}}}$, where $A_i$ denotes the observed outcome and $F_i$ denotes the forecasted outcome, the perfect forecasts of sun rise and financial crisis would result in the values of MSE and U to be 0 since $A_i$s are always equal to $F_i$s, indicating that both MSE and U cannot distinguish the difference between skill in perfect forecast of more common events such as sun rise and that of less frequent events such as financial crisis.

## 2.2. PSI for directional forecast of a single variable when a band of ± x of the forecasted outcome is considered

In order to evaluate the skill in direction-of-change forecast of a single variable when a band of ± *x*, where *x* > 0, of the forecasted outcome is considered, a few modifications in the 2 × 2 contingency table are made in order to recategorize the cell counts for *Hits* (*a*), *False alarms* (*b*), *Misses* (*c*), and *Correct rejections* (*d*). Let *Hits* (*a*) = Up/Up (within ± *x*), *False alarms* (*b*) = Up/Down + Down/Down (outside ± *x*), *Misses* (*c*) = Down/Up + Up/Up (outside ± *x*), and *Correct rejections* (*d*) = Down/Down (within ± *x*), respectively. The modified 2 × 2 contingency table of forecasted/observed outcomes of directional forecast of one variable, when a band of ± *x* of the forecasted outcome is introduced, is shown as Table 4.

Given the cell counts for *Hits* (*a*), *False alarms* (*b*), *Misses* (*c*), and *Correct rejections* (*d*) when a band of ± *x* of the forecasted outcome is considered as shown in Table 4, this study would like to clarify that the first priority would be given to the correct direction in that wrong-direction forecast would be counted as incorrect forecast (*False alarms* (*b*) or *Misses* (*c*)) even if the forecasted value falls within a band of ± *x* of the forecasted outcome. For example, if a band is set to be ± 0.5% of the forecasted outcome, the value of observed outcome of variable of interest at time *t* − 1 is 0%, the forecaster forecasts this variable at time *t* to be −0.25% ± 0.5%, and the observed outcome of this variable at time *t* turns out to be 0.25%, this would be counted as a wrong forecast since the forecaster forecasts this variable to go *Down* but, in fact, it does go *Up* despite its forecasted value falls within a band of ± 0.5% of the forecasted outcome.

To be counted as correct forecast when a band of ± *x* of the forecasted outcome is considered, the forecaster must forecast the direction correctly first. If the direction is forecasted correctly, then the forecasted value would be checked whether it falls within a band of ± *x* of the forecasted outcome. If the forecasted value falls within a band of ± *x* of the forecasted outcome, it would be counted as correct forecast (*Hits* (*a*) or *Correct rejections* (*d*)). If not, it would be counted as wrong forecast (*False alarms* (*b*) or *Misses* (*c*)). Note that even though the cell counts for *Hits* (*a*), *False alarms* (*b*), *Misses* (*c*), and *Correct rejections* (*d*) have been reclassified in order to account for the correct and incorrect forecasts when a band of ± *x* of the forecasted outcome is introduced, the definition of PSI(1) as shown in equation (1) can still be used to assess the skill of forecaster.

## 2.3. PSI for joint directional forecasts of multiple variables when a band of ± x of the forecasted outcome is considered

To jointly evaluate skill in directional forecasts of multiple variables, this study applies the method for constructing a composite index (I) devised by Sitthiyot and Holasut [19] which is shown as equation (3).

$$I = \sqrt{\frac{x^2 + y^2}{2}} \quad (3)$$

$$0 \leq I \leq 1$$

This composite index (I) takes values between 0 and 1 and comprises two variables which are *x* and *y*, where $0 \leq x \leq 1$ and $0 \leq y \leq 1$. Given that PSI(1) takes values between −1 and 1, in order to apply Sitthiyot and Holasut [19]'s method to construct PSI(N), a rescaling technique is used so that PSI(N) would take the values between −1 and 1. Let PSI(1)$_j$ denote the value of PSI(1) for the $j^{th}$ variable, where *j* = 1, 2, 3, ..., *m* and *m* = N. The rescaling could be done as shown in equation (4).

$$\text{PSI(N)} = \sqrt{\frac{(1 + \text{PSI(1)}_1)^2 + (1 + \text{PSI(1)}_2)^2 + \ldots + (1 + \text{PSI(1)}_m)^2}{N}} - 1 \quad (4)$$

$$-1 \leq \text{PSI(N)} \leq 1$$

when all PSI(1)$_j$s equal −1, PSI(N) equals −1 (the worst forecast skill). When all PSI(1)$_j$s equal 1, PSI(N) equals 1 (the best forecast skill). The reduced form of PSI for joint directional forecasts of N variables could be written as equation (5).

$$\text{PSI(N)} = \sqrt{\frac{\sum_{j=1}^{m} (1 + \text{PSI(1)}_j)^2}{N}} - 1 \quad (5)$$

**Table 4**
The modified 2x2 contingency table of forecasted/observed outcomes of directional forecast of one variable when a band of ± *x* of the forecasted outcome is considered.

| *Forecasted outcomes* | *Observed outcomes* | | | | Total |
|---|---|---|---|---|---|
| | Up | | Down | | |
| | within ± *x* | outside ± *x* | within ± *x* | outside ± *x* | |
| **Up** | *a* = Up/Up (within ± *x*) | Up/Up (outside +/− *x*) | *b* = Up/Down + | | *a* + *b* |
| **Down** | *c* = Down/Up + | | *d* = Down/Down (within ± *x*) | Down/Down (outside ± *x*) | *c* + *d* |
| **Total** | *a* + *c* | | *b* + *d* | | *n* = *a* + *b* + *c* + *d* |

$$-1 \leq \text{PSI(N)} \leq 1$$

To demonstrate how PSI(N) could be used to measure the skill of forecasters in practice, this study uses the data on Thailand growth of real gross domestic product (GDP growth) and inflation (Inf) forecasts made one year in advance by three organizations, namely, the Bank of Thailand (BOT), the Fiscal Policy Office (FPO), and the Office of the National Economic and Social Development Council (NESDC) as well as the actual data on Thailand annual GDP growth and Inf between 2001 and 2021 (total of 21 years) [26–28]. Note that the FPO's data on GDP growth and Inf forecasts are between 2003 and 2021 (total of 19 years). The data on GDP growth and Inf forecasts made one year in advance are collected from each organization. The actual data on annual GDP growth are collected from the NESDC while the actual data on annual Inf are collected from the BOT. All data used in this study are provided in Tables S1–S6 in Supplementary Materials. The summary statistics of the data on actual GDP growth and actual Inf as well as GDP growth and Inf forecasts made by the BOT, the FPO, and the NESDC are reported in Table 5.

Similar to the way in which PSI(1) measures skill in forecasting the direction-of-change of a single variable when a band of $\pm x$ of the forecasted outcome is considered as described earlier, the skill in directional forecasts of GDP growth and Inf one year ahead of each of the three organizations, namely, the BOT, the FPO, and the NESDC could be jointly evaluated by counting separately the forecasted/observed outcomes of direction-of-change forecasts of GDP growth and Inf when a band is considered, and then recording each of them in the modified $2 \times 2$ contingency table as shown in Table 4 accordingly. In this study, a band of $\pm 1$ standard deviation (SD) of the forecasted outcome is used. The bands of $\pm 1$ SD of the forecasted outcomes for GDP growth and Inf are calculated separately by using the data on the change in actual GDP growth and the change in actual Inf between 2001 and 2021 as shown in Tables S1–S6 in Supplementary Materials. Based on the summary statistics as reported in Table 5, the SD for the change in actual GDP growth is equal to 4.537% whereas the SD for the change in actual Inf is equal to 2.215%. In addition, a band of $\pm 0.5\%$ of the GDP growth and Inf forecasts is considered mainly because it is used in practice by the FPO [29].

The cell counts of forecasted/observed outcomes are then used to calculate PSI for directional forecasts of GDP growth (PSI(GDP)) and Inf (PSI(Inf)) separately for each organization. After obtaining the values of PSI(GDP) and those of PSI(Inf) for each of the three organizations, we can calculate PSI for joint directional forecasts of GDP growth and Inf (PSI(GDP&Inf)) for each of them by using the formula as shown in equation (5). Provided that there are two variables being forecasted at the same time, PSI(GDP&Inf) for each forecasting organization can be computed as shown in equation (6). Note that this study assigns equal weight to PSI(GDP) and PSI(Inf).

$$\text{PSI(GDP\&Inf)} = \sqrt{\frac{(1+\text{PSI(GDP)})^2 + (1+\text{PSI(Inf)})^2}{2}} - 1 \qquad (6)$$

$$-1 \leq \text{PSI(GDP\&Inf)} \leq 1$$

## 3. Results and discussion

### 3.1. The evaluation of skill in directional forecast when no band is considered

This study first reports the results of the evaluation of skills in directional forecasts when no band is imposed. The overall results, as shown in Table 6, suggest that the forecast skills of the BOT, the FPO, and the NESDC are far better than the random forecast since the values of PSI(GDP), PSI(Inf), and PSI(GDP&Inf) are greater than 0 in all cases. Considering the skill in directional forecast of GDP growth one year ahead, the BOT ranks 1st, followed by the FPO which ranks 2nd, and the NESDC which ranks 3rd with PSI(GDP)s = 0.718, 0.501, and 0.439, respectively. These results suggest that the skills in forecasting the direction-of-change of GDP growth of the BOT, the FPO, and the NESDC are 71.8%, 50.1%, and 43.9% better than the random forecast. For the skill in forecasting the direction-of-change of Inf one year in advance, the FPO ranks 1st with PSI(Inf) = 0.555, followed by the BOT which ranks 2nd with PSI(Inf) = 0.521. The NESDC ranks 3rd with PSI(Inf) = 0.332. These results indicate that the skills in forecasting Inf one year ahead of the FPO, the BOT, and the NESDC are 55.5%, 52.1%, and 33.2% better than the random forecast. The overall results show that, when evaluating the two variables separately and a band is not considered, the skills in forecasting the direction-of-change of GDP growth of the BOT and the NESDC are relatively better than that of Inf whereas the skill in forecasting the direction-of-change of Inf of the FPO is better than that of GDP growth.

**Table 5**
The summary statistics of the data on actual GDP growth and actual Inf as well as the data on GDP growth and Inf forecasts made by the BOT, the FPO, and the NESDC.

|  | Minimum | Maximum | Mean | Standard deviation | Number of observations (*n*) |
| --- | --- | --- | --- | --- | --- |
| **Actual** |  |  |  |  |  |
| GDP growth | −6.100 | 7.800 | 3.227 | 3.172 | 22 |
| Δ GDP growth | −8.500 | 10.100 | −0.152 | 4.537 | 21 |
| Inf | −0.900 | 5.500 | 1.856 | 1.784 | 22 |
| Δ Inf | −6.400 | 4.200 | −0.019 | 2.215 | 21 |
| **BOT** |  |  |  |  |  |
| GDP growth | 1.000 | 6.800 | 3.993 | 1.275 | 21 |
| Δ GDP growth | −4.250 | 3.300 | −0.028 | 1.659 | 20 |
| Inf | −0.500 | 4.250 | 1.843 | 1.345 | 21 |
| Δ Inf | −3.900 | 4.500 | −0.050 | 1.739 | 20 |
| **FPO** |  |  |  |  |  |
| GDP growth | −2.500 | 7.900 | 4.171 | 1.988 | 19 |
| Δ GDP growth | −8.000 | 7.000 | −0.128 | 2.770 | 18 |
| Inf | 0.300 | 4.500 | 2.295 | 1.297 | 19 |
| Δ Inf | −3.800 | 3.300 | −0.056 | 1.508 | 18 |
| **NESDC** |  |  |  |  |  |
| GDP growth | 2.000 | 7.500 | 4.038 | 1.148 | 21 |
| Δ GDP growth | −2.000 | 3.500 | −0.050 | 1.308 | 20 |
| Inf | 0.400 | 4.000 | 2.262 | 1.065 | 21 |
| Δ Inf | −1.900 | 1.300 | −0.038 | 0.738 | 20 |

For the joint evaluation of skills in directional forecasts made one year ahead of GDP growth and Inf, the results, as reported in Table 6, show that the BOT ranks 1st while the FPO ranks 2nd, followed by the NESDC which ranks 3rd with PSI(GDP&Inf)s = 0.623, 0.529, and 0.387, respectively. Provided that the benchmark score against which a forecast can be judged is the random forecast whose value is set to be 0%, the improvements over the random forecast by 62.3% for the BOT, 52.9%, for the FPO, and 38.7% for the NESDC suggest that, when no band is considered, all three organizations are somewhat skillful in forecasting one year in advance the direction-of-changes of GDP growth and Inf simultaneously.

*3.2. The evaluation of skill in directional forecast when a band of ± 1 SD of the forecasted outcome is considered*

As described in Materials and Methods, this study uses the data on the change in actual GDP growth and the change in actual Inf for calculating the SD for each variable. The calculated value of the SD for the change in actual GDP growth is equal to 4.537% whereas the calculated value of the SD for the change in actual Inf is equal to 2.215%. These SDs would be used as bands of ± 1 SD of the GDP growth and Inf forecasts accordingly. The results, as reported in Table 7, show that, by and large, there are no differences in forecast skills of the BOT, the FPO, and the NESDC compared to the cases when no band is imposed, except for the skill in forecasting Inf one year in advance and the skill in forecasting GDP growth and Inf simultaneously one year in advance of the NESDC whose values of PSI (Inf) and PSI(GDP&Inf) are equal to 0.236 and 0.341, respectively. These results suggest that the skill in forecasting Inf one year in advance and the skill in forecasting GDP growth and Inf simultaneously one year in advance of the NESDC are 23.6% and 34.1% better than the random forecast. Note that when no band is imposed, the skills of the NESDC in forecasting Inf one year ahead and in forecasting GDP growth and Inf simultaneously one year ahead are 33.2% and 38.7% better than the random forecast. However, when a band of ± 1 SD is imposed on the forecast outcome of Inf, the skill of the NESDC in forecasting Inf one year in advance is reduced to 23.6% better than the random forecast, resulting in the reduction in the NESDC's skill in forecasting GDP growth and Inf simultaneously one year in advance from 38.7% to 34.1% better than the random forecast.

*3.3. The evaluation of skill in directional forecast when a band of ± 0.5% of the forecasted outcome is considered*

While using a band of ± 1 SD where the SD is calculated based on the actual data is one reasonable way to determine the range of the forecasted outcome, in practice, the FPO uses a band of ± 0.5% of the forecast outcomes of GDP growth and Inf [29]. This study therefore follows the normal practice of the FPO by imposing a band of ± 0.5% on the values of GDP growth and Inf forecasts. The results are reported in Table 8. When a band of ± 0.5% of the forecasted outcome is introduced, the overall results of the evaluation of skills in directional forecast clearly indicate that the forecast performances of the BOT, the FPO, and the NESDC drop considerably relative to the cases when no band and a band of ± 1 SD of the forecasted outcome are considered. Focusing on the skill in directional forecast made one year in advance of GDP growth when a band of ± 0.5% of the forecasted outcome is considered, the BOT ranks 1st, followed by the FPO which ranks 2nd, and the NESDC which ranks 3rd with PSI(GDP)s = 0.164, 0.007, and 0.004, respectively. These results suggest that the skills in forecasting GDP growth one year in advance of the BOT, the FPO, and the NESDC are 16.4%, 0.7%, and 0.4% better than the random forecast. For the skill in forecasting one year ahead the direction-of-change of Inf when a band of ± 0.5% of the forecasted outcome is considered, the FPO ranks 1st with PSI(Inf) = 0.318. The BOT ranks 2nd with PSI(Inf) = 0.171, followed by the NESDC which ranks 3rd with PSI(Inf) = −0.055, indicating that the skills of FPO and the BOT in forecasting Inf one year ahead are 31.8% and 17.1% better than the random forecast whereas the skill of the NESDC in forecasting Inf one year ahead is 5.5% inferior to the random forecast.

**Table 6**
The evaluation of skills in directional forecasts of GDP growth, Inf, and both GDP growth and Inf one year in advance of the BOT, the FPO, and the NESDC when no band is considered.

| Forecasters | *Forecasted/Observed outcomes* | | | | | | | |
|---|---|---|---|---|---|---|---|---|
| | *Hits (a)* | *False alarms (b)* | *Misses (c)* | *Correct rejections (d)* | *n* | PSI(GDP) | PSI(Inf) | PSI(GDP&Inf) |
| | Up/Up | Up/Down | Down/Up | Down/Down | Total | [-1,1] | [-1,1] | [-1,1] |
| **BOT** | | | | | | | | |
| GDP growth | 9 | 2 | 1 | 9 | 21 | 0.718 | | |
| Inf | 9 | 3 | 2 | 7 | 21 | | 0.521 | |
| GDP growth&Inf | | | | | | | | 0.623 |
| **FPO** | | | | | | | | |
| GDP growth | 8 | 4 | 1 | 6 | 19 | 0.501 | | |
| Inf | 10 | 3 | 1 | 5 | 19 | | 0.555 | |
| GDP growth&Inf | | | | | | | | 0.529 |
| **NESDC** | | | | | | | | |
| GDP growth | 8 | 4 | 2 | 7 | 21 | 0.439 | | |
| Inf | 9 | 5 | 2 | 5 | 21 | | 0.332 | |
| GDP growth&Inf | | | | | | | | 0.387 |

**Table 7**
The evaluation of skills in directional forecasts of GDP growth, Inf, and both GDP growth and Inf one year in advance of the BOT, the FPO, and the NESDC when a band of ± 1 SD of the forecasted outcome is considered.

| Forecasters | Forecasted/Observed outcomes | | | | | | | PSI(GDP) | PSI(Inf) | PSI(GDP&Inf) |
| | *Hits (a)* | *False alarms (b)* | | *Misses (c)* | | *Correct rejections (d)* | n | | | |
| | Up/Up (within ± 1 SD) | Up/Down | Down/Down (outside ± 1 SD) | Down/Up | Up/Up (outside ± 1 SD) | Down/Down (within ± 1 SD) | Total | [-1,1] | [-1,1] | [-1,1] |
|---|---|---|---|---|---|---|---|---|---|---|
| **BOT** | | | | | | | | | | |
| GDP growth | 9 | 2 | 0 | 1 | 0 | 9 | 21 | 0.718 | | |
| Inf | 9 | 3 | 0 | 2 | 0 | 7 | 21 | | 0.521 | |
| GDP growth&Inf | | | | | | | | | | 0.623 |
| **FPO** | | | | | | | | | | |
| GDP growth | 8 | 4 | 0 | 1 | 0 | 6 | 19 | 0.501 | | |
| Inf | 10 | 3 | 0 | 1 | 0 | 5 | 19 | | 0.555 | |
| GDP growth&Inf | | | | | | | | | | 0.529 |
| **NESDC** | | | | | | | | | | |
| GDP growth | 8 | 4 | 0 | 2 | 0 | 7 | 21 | 0.439 | | |
| Inf | 8 | 5 | 1 | 2 | 1 | 4 | 21 | | 0.236 | |
| GDP growth&Inf | | | | | | | | | | 0.341 |

**Table 8**
The evaluation of skills in directional forecasts of GDP growth, Inf, and both GDP growth and Inf one year in advance of the BOT, the FPO, and the NESDC when a band of ± 0.5% of the forecasted outcome is considered.

| Forecasters | Forecasted/Observed outcomes | | | | | | | PSI (GDP) | PSI (Inf) | PSI (GDP&Inf) |
|---|---|---|---|---|---|---|---|---|---|---|
| | *Hits (a)* | *False alarms (b)* | | *Misses (c)* | | *Correct rejections (d)* | *n* | | | |
| | *Up/Up (within ± 0.5%)* | *Up/Down* | *Down/Down (outside ± 0.5%)* | *Down/Up* | *Up/Up (outside ± 0.5%)* | *Down/Down (within ± 0.5%)* | Total | [-1,1] | [-1,1] | [-1,1] |
| **BOT** | | | | | | | | | | |
| GDP growth | 5 | 2 | 7 | 1 | 4 | 2 | 21 | 0.164 | | |
| Inf | 4 | 3 | 3 | 2 | 5 | 4 | 21 | | 0.171 | |
| GDP growth&Inf | | | | | | | | | | 0.168 |
| **FPO** | | | | | | | | | | |
| GDP growth | 4 | 4 | 5 | 1 | 4 | 1 | 19 | 0.007 | | |
| Inf | 7 | 3 | 2 | 1 | 3 | 3 | 19 | | 0.318 | |
| GDP growth&Inf | | | | | | | | | | 0.173 |
| **NESDC** | | | | | | | | | | |
| GDP growth | 4 | 4 | 5 | 2 | 4 | 2 | 21 | 0.004 | | |
| Inf | 6 | 5 | 4 | 2 | 3 | 1 | 21 | | −0.055 | |
| GDP growth&Inf | | | | | | | | | | −0.025 |

Given that the reference score against which a forecast can be judged is the random forecast whose value is set to be 0%, these results indicate that the skill of the BOT in forecasting GDP growth one year in advance, when a band of ± 0.5% of the forecasted outcome is considered, is slightly better than the random forecast while the skills of the FPO and the NESDC in forecasting GDP growth one year in advance, when a band of ± 0.5% of the forecasted outcome is introduced, are indifferent from the random forecast. For the evaluation of skill in forecasting Inf one year ahead when a band of ± 0.5% of the forecasted outcome is considered, both the FPO and the BOT perform better than the random forecast while the NESDC performs no better than the random forecast.

Considering the joint evaluation of skills in directional forecasts of GDP growth and Inf one year in advance when a band of ± 0.5% of the forecasted outcome is considered, the results, as reported in Table 8, show that the FPO ranks 1st while the BOT ranks 2nd, followed by the NESDC which ranks 3rd with PSI(GDP&Inf)s = 0.173, 0.168, and −0.025, respectively. These results suggest that the performances of the FPO and the BOT in forecasting the direction-of-changes of GDP growth and Inf simultaneously one year in advance are slightly better than the random forecast since the skills of the FPO and the BOT in forecasting GDP growth and Inf simultaneously one year in advance are 17.3% and 16.8% better than the random forecast while the skill in forecasting both variables simultaneously one year in advance of the NESDC is 2.5% worse than the random forecast.

## 4. Conclusions and remarks

Almost always forecasts for key macroeconomic variables are made simultaneously by the same organizations, presented together, and used together in policy analyses and decision-makings [3], it is therefore important to know whether the forecasters have enough skills in forecasting the direction-of-changes of multiple variables at the same time. This study introduces a method for joint evaluation of skill in directional forecasts of multiple variables. Our method is simple to use and does not rely on complicated assumptions required by the conventional statistical methods for measuring directional forecast accuracy. To demonstrate how the method can be used to evaluate the skills of forecasters in practice, the forecast and actual data on Thailand GDP growth and Inf between 2001 and 2021 from three organizations, namely, the BOT, the FPO, and the NESDC, are used. The results show that, when no band and a band of ± 1 SD of the forecasted outcome are considered, all three organizations are somewhat skillful in forecasting the direction-of-changes of GDP growth and Inf one year in advance when assessing both variables separately and simultaneously. However, when a band of ± 0.5% of the forecasted outcome is considered, the skills in directional forecasts of GDP growth and Inf one year in advance of the BOT are slightly better than intelligent guess work when evaluating the two variables separately and simultaneously. The FPO has skill in forecasting the direction of Inf one year ahead but not GDP growth. However, when assessing both variables simultaneously, the skills of the FPO in forecasting GDP growth and Inf one year in advance are little better than chance. For the NESDC, its skills in forecasting GDP growth and Inf one year ahead are no better than the random forecast when assessing the two variables separately and simultaneously. Based on these results, it can be concluded that the BOT, the FPO, and the NESDC, are somewhat reliable in forecasting the direction-of-changes of GDP growth and Inf one year in advance when no band and a band of ± 1 SD of the forecasted outcome are considered. However, when a band of ± 0.5% of the forecasted outcome is imposed, the skills of these three organizations in forecasting the direction-of-changes of GDP growth and Inf one year ahead are, at best, little better than the random forecast.

Provided that a good forecast is of importance for scientific, economic, and administrative purposes [30], we hope that our method could be used to improve the skills of forecasters who build models in order to forecast not only direction-of-changes of economic variables but also binary outcomes/events in other scientific disciplines as well. However, it should be noted that our method measures forecast skill only. There are other scalar aspects of forecast quality such as accuracy, bias, reliability, resolution, discrimination, and sharpness [1] which could be used in combination with the PSI(N) method in order to assess the overall forecast performance. Last but not least, we do acknowledge that forecast errors do matter. However, from our viewpoint, the direction-of-change should be given the first priority. Once we would be able to correctly forecast the direction-of-change of a variable on a consistent basis, the size of the forecast errors could later be adjusted.

**Author contribution statement**

**Data availability statement**

All data generated and/or analyzed during this study are included in this article and its Supplementary Materials.

**Funding statement**

This research did not receive any specific grant from funding agencies in the public, commercial, or not-for-profit sectors.

**Declaration of competing interest**

The authors declare that they have no known competing financial interests or personal relationships that could have appeared to influence the work reported in this paper.

**Acknowledgements**

The authors are grateful to Dr. Suradit Holasut and two reviewers for guidance and comments.

**Appendix A. Supplementary data**

Supplementary data to this article can be found online at https://doi.org/10.1016/j.heliyon.2023.e19729.

**References**

[1] D.S. Wilks, Statistical Methods in the Atmospheric Sciences, third ed., Academic Press, San Diego, 2011, pp. 305–316.
[2] H.O. Stekler, Perspectives on evaluating macroeconomic forecasts, in: M.L. Higgins (Ed.), Advances in Economic Forecasting, W.E. Upjohn Institute for Employment Research, Kalamazoo, 2011, pp. 105–148.
[3] T.M. Sinclair, H.O. Stekler, L. Kitzinger, Directional forecasts of GDP and inflation: a joint evaluation with an application to Federal Reserve predictions, Appl. Econ. 42 (2010) 2289–2297, https://doi.org/10.1080/00036840701857978.
[4] M.H. Pesaran, A. Timmermann, A simple nonparametric test of predictive performance, J. Bus. Econ. Stat. 10 (1992) 461–465.
[5] T. Sitthiyot, K. Holasut, On the evaluation of skill in binary forecast, Thailand World Econ 40 (2022) 33–54. https://so05.tci-thaijo.org/index.php/TER/article/view/261138.
[6] Z. Md Faisal, S.S. Monira, H. Hirose, DF-ReaL2Boost: a hybrid decision forest with Real L2Boost decision stumps, in: F.L. Gaol (Ed.), Recent Progress in Data Engineering and Internet Technology, vol. 1, Springer, New York, 2013, pp. 47–53.
[7] L.R. Dice, Measures of the amount of ecologic association between species, Ecology 26 (1945) 297–302.
[8] K. Lahiri, L. Yang, Forecasting binary outcomes, in: G. Elliott, A. Timmermann (Eds.), Handbook of Economic Forecasting, vol. 2, North-Holland, New York, 2013, pp. 1025–1106, part B.
[9] R. So, A. Teakles, J. Baik, R. Vingarzan, K. Jones, Development of visibility forecasting modeling framework for the lower fraser valley of British columbia using Canada's regional air quality deterministic prediction system, J. Air Waste Manag. Assoc. 68 (2018) 446–462.
[10] C.W.J. Granger, M.H. Pesaran, Economic and statistical measures of forecast accuracy, J. Forecast. 19 (2000) 537–560.
[11] L. Foresti, M. Reyniers, A. Seed, L. Delobbe, Development and verification of a real-time stochastic precipitation nowcasting system for urban hydrology in Belgium, Hydrol. Earth Syst. Sci. 20 (2016) 505–527.
[12] A. McGovern, D.J. Gagne II, J.K. Williams, R.A. Brown, J.B. Basara, Enhancing understanding and improving prediction of severe weather through spatiotemporal relational learning, Mach. Learn. 95 (2014) 27–50.
[13] E. Camporeale, The Challenge of Machine Learning in Space Weather: Nowcasting and Forecasting, 17, Space Weather, 2019, pp. 1166–1207.
[14] W. Briggs, D. Ruppert, Assessing the skill of yes/no predictions, Biometrics 61 (2005) 799–807.
[15] H. Halide, Implementing predictive models for domestic decision-making against dengue haemorrhagic fever epidemics, Dengue Bull. 33 (2009) 1–10.
[16] A. Manzato, I.T. Jolliffe, Behaviour of verification measures for deterministic binary forecasts with respect to random changes and thresholding, Q. J. R. Meteorol. Soc. 143 (2017) 1903–1915.
[17] J.R. Holliday, J.B. Rundle, D.L. Turcotte, Earthquake forecasting and verification, in: R.A. Meyers (Ed.), Encyclopedia of Complexity and Systems Science, Springer, Berlin, 2009, pp. 2438–2449.
[18] Y. Kubo, M. Den, M. Ishii, Verification of operational solar flare forecast: case of regional warning center Japan, J. Space Weather Space Clim. 7 (A20) (2017) 1–16, https://doi.org/10.1051/swsc/2017018.
[19] T. Sitthiyot, K. Holasut, A simple method for measuring inequality, Palgrave Commun 6 (2020) 112, https://doi.org/10.1057/s41599-020-0484-6.
[20] C.S. Peirce, The numerical measure of the success of predictions, Science 4 (1884) 453–454.
[21] G.U. Yule, On the methods of measuring association between two attributes, J. Roy. Stat. Soc. 75 (1912) 576–642.
[22] P. Heidke, Berechnung des erfolges und der güte der windstärkvorhersagen im sturmwarnungsdienst, Geografika Annaler 8 (1926) 301–349.
[23] H.H. Clayton, Rating weather forecasts, Bull. Am. Meteorol. Soc. 15 (1934) 279–283.
[25] H. Theil, Applied Economic Forecasting, North-Holland, Amsterdam, 1966, p. 28.
[26] Bank of Thailand, Monetary policy report. https://www.bot.or.th/English/MonetaryPolicy/MonetPolicyComittee/MPR/Pages/default.aspx, 2022.
[27] Fiscal Policy Office, Thailand Economic Outlook, 2022. http://www.fpo.go.th/main/Economic-report/Thailand-Economic-Projections.aspx.
[28] Office of the National Economic and Social Development Council, Thai Economic Performance, 2022. https://www.nesdc.go.th/main.php?filename=macroeconomics.
[29] Fiscal Policy Office, Thailand's Economic Outlook 2021, 2021. https://www.fpo.go.th/main/Economic-report/Thailand-Economic-Projections/14444.aspx.
[30] F. Woodcock, The evaluation of yes/no forecasts for scientific and administrative purposes, Mon. Weather Rev. 104 (1976) 1209–1214.